\documentclass[12pt]{article}
\usepackage{amsmath}
\usepackage{amssymb}
\usepackage{epsfig}
\usepackage{euscript}
\usepackage{fancybox}
\usepackage{color}

\def\mathswitchr#1{\relax\ifmmode{\mathrm{#1}}\else$\mathrm{#1}$\fi}

\newcommand {\pslash}{\hbox{$\not\hbox{\kern-2.3pt $p$}$}}

\usepackage{cite}

\usepackage{epic}

\def\alf1{ {\alpha\over\pi} }

\begin{document}
\begin{titlepage}
\begin{flushright}
{\bf BU-HEPP-10-03 }\\
{\bf June, 2010}\\
\end{flushright}
 
\begin{center}
{\Large ``Low'' Energy GUTs $^{\dagger}$
}
\end{center}

\vspace{2mm}
\begin{center}
{\bf   B.F.L. Ward}\\
\vspace{2mm}
{\em Department of Physics,\\
 Baylor University, Waco, Texas, 76798-7316, USA}\\
\end{center}

\vspace{5mm}
\begin{center}
{\bf   Abstract}
\end{center}
We introduce a new
approach to the subject of grand unification
which allows the GUT scale to be small, $\lesssim 200$TeV, so that it is within
the reach of {\em conceivable} laboratory accelerated colliding beam devices. Central to the approach is a novel abstraction of the heterotic string symmetry
group physics ideas to
render baryon number violating effects small enough to
have escaped detection to date.
\vspace{10mm}
\vspace{10mm}
\renewcommand{\baselinestretch}{0.1}
\footnoterule
\noindent
{\footnotesize
\begin{itemize}
\item[${\dagger}$]
Work partly supported 
by NATO Grant PST.CLG.980342.
\end{itemize}
}

\end{titlepage}

\def\Kmax{K_{\rm max}}\def\ieps{{i\epsilon}}\def\rQCD{{\rm QCD}}
\renewcommand{\theequation}{\arabic{equation}}
\font\fortssbx=cmssbx10 scaled \magstep2
\renewcommand\thepage{}
\parskip.1truein\parindent=20pt\pagenumbering{arabic}\par
\indent The structure 
of the Standard Model(SM)~\cite{sm1,qcd}, in view of its success, 
leads naturally to the suggestion that
all forces associated with the gauge interactions therein may
be unified into a single gauge principle associated
with a larger group ${\cal G}$ which contains 
the SM gauge group $SU(2)_{L}\times U(1)_Y\times SU(3)^c$
as a subgroup, where we use a standard notation for the SM gauge group.
Originally introduced in the modern context in
Refs.~\cite{pati-salam,geor-glash}, this idea continues to
be a fashionable area of investigation today, where 
approaches which unify the SM gauge forces with that of
quantum gravity are now very much in vogue via the superstring
theory~\cite{gsw,jp} and its various low energy 
reductions and
morphisms~\cite{jp}. In what follows here, we focus only on the unification
of the SM gauge forces themselves, candidates for 
which we call as usual GUTs, so that we leave aside 
any possible unification with quantum gravity until a later
study~\cite{elswh}. 
\par
We admit that a part of our motivation is the recent
progress in the approaches to the Einstein-Hilbert theory for quantum gravity
in Refs.~\cite{asympsfty,bw1,kreimer,lpqg}
in which improved treatments of perturbation theory via resummation methods,
the asymptotic safety approach, the resummed quantum
gravity approach or the Hopf-algebraic Dyson-Schwinger
equation renormalization theory approach, 
and the introduction of
an underlying loop-space at Planck scales, loop quantum gravity,
all support the view that the apparently bad unrenormalizable
behavior of the Einstein-Hilbert theory may be cured by 
the dynamical interactions or
modifications within the theory itself, as first anticipated by Weinberg~\cite{asympsfty}. 
In what follows, we explore the suggestion, which follows from such progress, 
that the unification of all other forces 
can be a separate problem
from the
problem of treating the apparently
bad UV behavior of quantum gravity.\par 

Our objective is to formulate GUTs so that they are accessible
to very high energy colliding beam devices such as the VLHC,
which has been discussed elsewhere~\cite{vlhc} with cms energies
in the 100-200TeV regime. We show in what follows that we can achieve such
GUTs that satisfy the usual requirements: no anomalies,
unified SM couplings, baryon stability, absence/suppression 
of other unwanted transitions and naturalness requirements (this may
just mean N=1 susy here~\cite{wittn}). Here, we add the new condition
that the theory will live in 4-dimensional Minkowski space.
We call this our {\em known physical reality condition}. 
The most demanding requirement will be
seen to be baryon stability.\par

To illustrate why the most difficult aspect of a 
GUT with a (several) hundred TeV
unification scale is the issue of baryon number stability we
note that the proton
must be stable to $\sim 10^{29-33}$yrs, depending on the mode.
Standard methods can be used to show that
the natural lifetime
for physics with a $100$TeV scale for a dimension 6 transition 
in a state with the size and mass of the proton is
$\sim 0.01$yr for example. Clearly, some new mechanism is needed
to suppress the proton decay process here.\par

In proceeding to isolate such a mechanism, we will use 
what is sometimes called a radically conservative 
approach - we will try to rely on well-tested ideas used in a novel way.
In this way we may hope to avoid 
moving the GUT scale to $\sim 10^{13}$TeV as it is usually done~\cite{gut1},
or invoking hitherto unknown phenomena, such 
as extra dimensions~\cite{kdgut,gut1},
etc. We notice that the fundamental structure of a GUT theory has it
organized by gauge sector, by family sector and by Higgs
sector for spontaneous symmetry breaking. We turn now to the
family and gauge sectors. Let us also note that, in effecting this discussion, 
we present here a different realization of the basic ideas 
we already introduced in Ref.~\cite{bw2}. Only experiment can tell us
which realization is used by Nature.\par

Specifically, the ${\bf 10+\bar{5}}$ of 
$SU(5)$ was advocated in Ref.~\cite{geor-glash} and shown
to accommodate the SM family with a massless neutrino. With the recent
advent of neutrino masses~\cite{neut1,neut2}, we must
extend this fifteen dimensional representation to
a sixteen dimensional representation. We choose 
to use the ${\bf 16}$ of $SO(10)$~\cite{gross},
as it decomposes as  ${\bf 10+\bar{5}\bf+1}$ under an inclusion of $SU(5)$
into $SO(10)$. From the heterotic 
string formalism~\cite{gsw,jp}(we view here modern string theory 
as an extension of quantum field theory which can be used to abstract
dynamical relationships which would hold in the real world even if
the string theory itself is in detail only an approximate, 
mathematically consistent 
treatment of that reality, just as the old strong interaction string theory~\cite{schwz1} could be used to abstract properties
of QCD such as Regge trajectories even before QCD was discovered)
we know that in the only known and accepted unification of the SM and gravity,
the gauge group $E_8\times E_8$ is singled-out
when all known dualities~\cite{jp} are taken into account to relate equivalent superstring theories. A standard breakdown of this
symmetry to the SM gauge group and family structure is as follows~\cite{jp}:
\begin{equation}
\begin{split}
E_8&\rightarrow SU(3)\times E_6 \rightarrow SU(3)\times SO(10)\times U'(1)\nonumber\\ 
&\rightarrow SU(3)\times SU(5)\times U''(1)\times U'(1)\nonumber\\
&\rightarrow  SU(3)\times SU(3)^c\times SU(2)_L\times U(1)_Y\times U''(1)\times U'(1)
\end{split}
\end{equation}
where the SM gauge group is now called out as $SU(3)^c\times SU(2)_L\times U(1)_Y$. It can be shown that the ${\bf 248}$ of $E_8$ then splits
under this breaking into $({\bf 8,1})+({\bf 1,78})+({\bf 3,27})+(\bar{\bf{3}},
\overline{\bf{27}})$
under $SU(3)\times E_6$ and that each ${\bf{27}}$ under $E_6$ contains exactly
one SM family 16-plet with 11 other states that are paired with their anti-particles in helicity via real representations 
so that they would be expected to become massive at the GUT scale.
Let us consider that we have succeeded with the heterotic string breaking
scenario to get 6 families~\cite{gut1} 
under the first $E_8$ factor, $E_{8a}$, in the $E_8\times E_8$
gauge group. They are singlets under the second $E_8\equiv E_{8b}$.
We take the first 3 families to be those with the known light leptons
and the remaining 3 families to be those with the known light quarks.
The quarks in the families with the known light leptons are at a scale $M_{QL}$
that is beyond current experimental limits on new quarks; the leptons in 
the families with the known light quarks are at a scale $M_{LL}$ 
that is beyond the current experimental limits on heavy leptons.
We now repeat the same pattern of breaking for the second factor $E_{8b}$
as well
and we leave open the issue of observable families under this $E_{8b}$, as
they may exist in principle as well. The scales $M_{QL},M_{LL}$ are bounded
by the grand unified theory (GUT) scale $M_{GUT}$.
This scenario stops baryon instability: the proton can not decay 
because the leptons
to which it could transform via (leptoquark) bosons are all at too
high a scale. The extra heavy quarks and leptons just introduced here may of course appear already at the LHC.
\par
The ordinary electroweak and strong interaction
gauge bosons are now an unknown mixture
of the two copies of such bosons from the two $E_8's$
associated to heterotic string theory\footnote{If one wants to avoid any reference to superstring theory, one can just postulate our symmetry and families as needed, obviously; we leave this to the discretion of the reader.}:
when we break the two $E_8$'s  each to a product group
$SU(3)\times E_6$ and then subsequently break each of
the two $E_6$'s to get two copies 
of ${SU(3)}^c\times SU(2)_L\times U(1)_Y$, for
the initially massless gauge bosons for $SU(3)^c_i\times SU(2)_{Li}\times U(1)_{Yi}\in E_{8i}$,~$G^a_i,~a=1,\cdots,8,~A^{i'}_i,~i'=1,\cdots,3,~B_i$,\ $i=1,2$, in a standard notation, we assume a further breaking at the 
GUT scale so that the following linear combinations are massless at the GUT scale $M_{GUT}$ while the orthogonal linear combinations acquire masses ${\cal O}(M_{GUT})$ --
\begin{equation}
\label{fgauge1}
\begin{split}
A_f^{i'}=\sum_{i=1}^{2}\eta_{2i} A^{i'}_i\\
B_f= \sum_{i=1}^{2}\eta_{1i} B_i .
\end{split}
\end{equation}
The mixing coefficients $\{\eta_{aj}\}$ satisfy
$$\sum_{i=1}^{2}\eta_{ai}^2=1,\; a=1,2$$.\par
For the strong interaction, we take the minimal view that 
the quarks in each of the families from the two $E_8$'s are confined.
We use discrete symmetry to set the two strong interaction
gauge couplings to be equal at the GUT scale. This
means that for the known quarks we have gluons $G^a_1$. Of course,
experiments may ultimately force us to break the as yet unseen
color group. This is straightforward to do following Ref.~\cite{LFLi}.\par
For the low energy EW bosons, we have some 
freedom in (\ref{fgauge1}). We note the following values~\cite{siggi,pdg08} 
of the known
gauge couplings at scale $M_Z$: 
\begin{equation}
\label{fgauge2}
\begin{split}
&\alpha_s(M_Z)|_{\overline{MS}}= 0.1184\pm 0.0007\\
&\alpha_W(M_Z)|_{\overline{MS}}= 0.033812\pm 0.000021\\
&\alpha_{EM}(M_Z)|_{\overline{MS}}= 0.00781708\pm0.00000098
\end{split}
\end{equation}
It is well-known~\cite{gqw} that the factor of almost 4 between $\alpha_s(M_Z)$ and $\alpha_W(M_Z)$ and between $\alpha_W(M_Z)$ and $\alpha_{EM}(M_Z)$ when the respective unified values
are 1 and 2.67 require $M_{GUT}\sim 10^{13}-10^{12}$TeV. Here, with the use of the $\{\eta_{kj}\}$ we can absorb most of the discrepancy between the unification and
observed values of the coupling ratios so that the GUT scale is not beyond
current technology for accelerated colliding beam devices.\par
More precisely, we can set 
\begin{equation}
\label{fgauge3}
\begin{split}
\eta_{21}\cong \frac{1}{\sqrt{2.000}}\\
\eta_{11}\cong  \frac{1}{\sqrt{3.260}}
\end{split}
\end{equation} 
and this will leave a ``small'' amount of evolution do be done
between the scale $M_Z$ and $M_{GUT}$.\par
Indeed, with the choices in (\ref{fgauge3}), and the use of the one-loop 
beta functions~\cite{qcd}, if we use continuity of the gauge coupling constants at mass thresholds with one such threshold at $m_H\cong 120$GeV and a second one at $m_t=171.2$GeV for definiteness to illustrate our approach, then the GUT scale can be easily evaluated to be $M_{GUT}\cong 136$TeV, as advertised. For, we get,
\begin{equation}
\label{fgauge4}
b^{U(1)_Y}_{0} = \frac{1}{12\pi^2}\begin{cases}4.385&,\;M_Z\le \mu\le m_H\cong 120\text{GeV}\\ 4.417&, \; m_H< \mu \le m_t\\ 
5.125&, \; m_t < \mu \le M_{\text{GUT}}
\end{cases}
\end{equation}
from the standard formula~\cite{qcd} 
\begin{equation}
\label{fgauge5}
b^{U(1)_Y}_{0}=\frac{1}{12\pi^2}\left(\sum_j n_j\left(\frac{Y_j}{2}\right)^2\right)
\end{equation}
where $b^{U(1)_Y}_0$ is the coefficient of $g'^3$ in the beta function for the
$U(1)_Y$ coupling constant 
$g'$ in the $SU(2)_{L}\times U(1)_Y$ EW theory of 
Glashow, Salam and Weinberg~\cite{sm1}, 
$n_j$ is the effective number of Dirac fermion degrees of freedom,
i.e., a left-handed Dirac fermion counts as $\frac{1}{2}$, a complex scalar
counts as $\frac{1}{4}$, and so on. Similarly, for the QCD and $SU(2)_{L}$
theories, we get the analogous
\begin{equation}
\label{fgauge6a}
b^{SU(2)_{L}}_{0} = \frac{-1}{16\pi^2}\begin{cases}3.708&,\;M_Z\le \mu\le m_H\cong 120\text{GeV}\\ 3.667&, \; m_H< \mu \le m_t\\ 
3.167&, \; m_t < \mu \le M_{\text{GUT}}
\end{cases}
\end{equation}
\begin{equation}
\label{fgauge6b}
b^{QCD}_{0} = \frac{-1}{16\pi^2}\begin{cases}7.667&,\;M_Z\le \mu\le m_t\\ 7&, \; m_t < \mu \le M_{\text{GUT}}
\end{cases}
\end{equation}
from the standard formula~\cite{qcd} 
\begin{equation}
\label{fgauge7}
b^{\cal H}_{0}=\frac{-1}{16\pi^2}\left(\frac{11}{3}C_2({\cal H})-\frac{4}{3}\sum_j n_jT(R_j)\right)
\end{equation}
where $T(R_j)$ sets the normalization of the generators $\{\tau^{R_j}_a\}$ of 
the group ${\cal H}$ in the representation $R_j$ via $\text{tr}\tau^{R_j}_a\tau^{R_j}_b=T(R_j)\delta_{ab}$ where $\delta_{ab}$ is the Kronecker delta and
$C_2({\cal H})$ is the quadratic Casimir invariant eigenvalue for the adjoined representaion of ${\cal H}$.
These results 
(\ref{fgauge4},\ref{fgauge5},\ref{fgauge6a},\ref{fgauge6b},\ref{fgauge7})
together with the standard one-loop solution~\cite{qcd}
\begin{equation}
\label{fgauge8}
g^2_{\cal H}(\mu)=\frac{g^2_{\cal H}(\mu_0)}{1-2b^{\cal H}_0 g^2_{\cal H}(\mu_0)\ln(\mu/\mu_0)}
\end{equation} 
allow us to compute the value $M_{GUT}\cong 136$TeV for the values of 
$\eta_{ij}$ given in (\ref{fgauge3}). Here, we use standard notation
that $g^2_{\cal H}(\mu)$ is the squared running coupling constant at scale $\mu$
for ${\cal H}=U(1)_Y, \; SU(2)_{L}, \; QCD\equiv SU(3)^c$. 
\par
For illustration we have chosen the value of $136$TeV for the unification 
scale.In principle any value between the TeV scale and the Planck scale 
is allowed in our approach and wait for 
experiment to tell us what the true value is.\par
We sum up with the following observation, already made in Ref.~\cite{bw2}:
instead of the traditional ``desert''~\cite{geor-glash,gqw} between the TeV scale and the GUT scale, we propose here a ``green pasture''.\par

\end{document}